\documentstyle[preprint,aps]{revtex}

\begin{document}

\draft

\preprint{July,2000}

\title{Yang-Mills Instantons in the Gravitational Instanton Backgrounds}

\author{Hongsu Kim\footnote{e-mail : hongsu@hepth.hanyang.ac.kr} and
Yongsung Yoon\footnote{e-mail : cem@hepth.hanyang.ac.kr}}

\address{Department of Physics \\
Hanyang University, Seoul, 133-791, KOREA}

%\date{January, 2000}

\maketitle

\begin{abstract}
The simplest and the most straightforward new algorithm for generating
solutions to (anti) self-dual Yang-Mills (YM) equation in the
typical gravitational instanton backgrounds is proposed.
When applied to the Taub-NUT and the Eguchi-Hanson metrics, the two
best-known gravitational instantons, the solutions turn out to
be the rather exotic type of instanton configurations
carrying finite YM action but generally fractional topological charge values.

\end{abstract}


\narrowtext
%\twocolumn

 Well below the Planck scale, the strength of gravity is
 negligibly small relative to those of particle physics
 interactions described by non-abelian gauge theories.
 Nevertheless, as far as the topological aspect is concerned,
 gravity may have marked effects even at the level of elementary
 particle physics. Namely, the non-trivial topology of the
 gravitational field may play a role crucial enough to dictate the
 topological properties of, say, SU(2) Yang-Mills (YM) gauge
 field [1] as has been pointed out long ago [2]. Being an issue
 of great physical interest and importance, quite a few serious study
 along this line have appeared in the literature but they were
 restricted to the background gravitational field with high degree
 of isometry such as the Euclideanized Schwarzschild
 geometry [2] or the Euclidean de Sitter space [3]. Even the works
 involving more general background spacetimes including gravitational
 instantons (GI) were mainly confined to the case of asymptotically-
 locally-Euclidean (ALE) spaces which is one particular such GI and
 employed rather indirect and mathmatically-oriented solution generating
 methods such as the ADHM construction [11]. Here in this work we would
 like to propose a ``simply physical'' and hence perhaps the most direct
 algorithm for generating the YM instanton solutions in all species
 of known GI. And the essence of this method lies in writing the (anti)
 self-dual YM equation by employing truly relevant ans\H{a}tz for the
 YM gauge connection and then directly solving it.
 To demonstrate how simple in method and powerful in applicability
 it is, we then apply this algorithm to the case of the Taub-NUT
 and the Eguchi-Hanson metrics, the two best-known GI. In particular,
 the actual YM instanton solution in the background of Taub-NUT metric
 (which is asymptotically-locally-flat (ALF) rather than ALE) is
 constructed for the first time in this work although its existence has
 been anticipated long ago in [2].  Interestingly, the solutions to
 (anti) self-dual YM equation turn out to be the rather exotic type of
 instanton configurations which are everywhere non-singular having {\it finite}
 YM action but sharing some features with meron solutions [9] such as
 their typical structure and generally {\it fractional} topological  charge values
 carried by them. Namely, the YM instanton solution that we shall discuss in
 the background of GI in this work exhibit characteristics which are mixture
 of those of typical instanton and typical meron. This seems remarkable
 since it is well-known that in flat spacetime, meron does not solve the
 1st order (anti) self-dual equation although it does the 2nd order YM
 field equation and is singular at its center and has divergent action. \\
 In the loose sense, GI may be
 defined as a positive-definite metrics $g_{\mu\nu}$ on a complete
 and non-singular manifold satisfying the Euclidean Einstein
 equations and hence constituting the stationary points of the
 gravity action in Euclidean path integral for quantum gravity.
 But in the stricter sense [4], they are the metric solutions to the
 Euclidean Einstein equations having (anti) self-dual Riemann
 tensor
\begin{eqnarray}
\tilde{R}_{abcd} = {1\over 2}\epsilon_{ab}^{~~ef} R_{efcd} = \pm
R_{abcd}
\end{eqnarray}
(say, with indices written in non-coordinate orthonormal basis)
and include only two families of solutions in a rigorous sense ;
the Taub-NUT metric [5] and the Eguchi-Hanson instanton [6].
In the loose sense, however, there are several solutions to
Euclidean Einstein equations that can fall into the category of GI.
Thus we begin with the action governing our
system, i.e., the Einstein-Yang-Mills (EYM) theory given by
\begin{eqnarray}
I_{EYM} = \int_{M} d^4x\sqrt{g}\left[{-1\over 16\pi}R + {1\over
4g^2_{c}}F^{a}_{\mu\nu}F^{a\mu\nu}\right] - \int_{\partial M}
d^3x\sqrt{h}{1\over 8\pi}K
\end{eqnarray}
where $F^{a}_{\mu\nu}$ is the field strength of the YM gauge field
$A^{a}_{\mu}$ with $a=1,2,3$ being the SU(2) group index and
$g_{c}$ being the gauge coupling constant. The Gibbons-Hawking
term on the boundary $\partial M$ of the manifold $M$ is also
added and $h$ is the metric induced on $\partial M$ and $K$ is the
trace of the second fundamental form on $\partial M$. Then by
extremizing this action with respect to the metric $g_{\mu\nu}$
and the YM gauge field $A^{a}_{\mu}$, one gets the following
classical field equations respectively
\begin{eqnarray}
&&R_{\mu\nu} - {1\over 2}g_{\mu\nu}R  =
8\pi T_{\mu\nu}, \nonumber \\
&&T_{\mu\nu} = {1\over
g^2_{c}} \left[F^{a}_{\mu\alpha}F_{\nu}^{a\alpha} - {1\over 4}
g_{\mu\nu}(F^{a}_{\alpha\beta}F^{a\alpha\beta})\right], \\
&&D_{\mu}\left[\sqrt{g}F^{a\mu\nu}\right] = 0,
~~~D_{\mu}\left[\sqrt{g}\tilde{F}^{a\mu\nu}\right] = 0 \nonumber
\end{eqnarray}
where we added Bianchi identity in the last line and
$F^{a}_{\mu\nu} =
\partial_{\mu}A^{a}_{\nu}-\partial_{\nu}A^{a}_{\mu}+\epsilon^{abc}
A^{b}_{\mu}A^{c}_{\nu}$, $D^{ac}_{\mu} = \partial_{\mu}\delta^{ac}
+\epsilon^{abc}A^{b}_{\mu}$ and $A_{\mu}=A^{a}_{\mu}(-iT^{a})$,
$F_{\mu\nu}=F^{a}_{\mu\nu}(-iT^{a})$ with $T^{a}=\tau^{a}/2$
($a=1,2,3$) being the SU(2) generators and finally
$\tilde{F}_{\mu\nu} = {1\over 2}
\epsilon_{\mu\nu}^{~~\alpha\beta}F_{\alpha\beta}$ is the (Hodge) dual
of the field strength tensor. We now seek solutions ($g_{\mu\nu}$,
$A^{a}_{\mu}$) of the coupled EYM equations given above in
Euclidean signature obeying the (anti) self-dual equation in the
YM sector
\begin{eqnarray}
F^{\mu\nu} = g^{\mu\lambda}g^{\nu\sigma}F_{\lambda\sigma} = \pm
{1\over 2} \epsilon_{c}^{\mu\nu\alpha\beta}F_{\alpha\beta}
\end{eqnarray}
where
$\epsilon_{c}^{\mu\nu\alpha\beta}=\epsilon^{\mu\nu\alpha\beta}/\sqrt{g}$
is the curved spacetime version of totally antisymmetric tensor.
As was noted in [2,3], in Euclidean signature, the YM
energy-momentum tensor vanishes identically for YM fields
satisfying this (anti) self-duality condition. This point is of
central importance and can be illustrated briefly as follows.
Under the Hodge dual transformation, $F^{a}_{\mu\nu} \rightarrow
\tilde{F}^{a}_{\mu\nu}$, the YM energy-momentum tensor
$T_{\mu\nu}$ given in eq.(3) above is invariant normally in
Lorentzian signature. In Euclidean signature, however, its sign
flips, i.e., $\tilde{T}_{\mu\nu} = - T_{\mu\nu}$. As a result, for
YM fields satisfying the (anti) self-dual equation in Euclidean
signature such as the instanton solution, $F^{a}_{\mu\nu} = \pm
\tilde{F}^{a}_{\mu\nu}$, it follows that $T_{\mu\nu} =
-\tilde{T}_{\mu\nu} = -T_{\mu\nu}$, namely the YM energy-momentum
tensor vanishes identically, $T_{\mu\nu}=0$. This, then, indicates
that the YM field now does not disturb the geometry while the
geometry still does have effects on the YM field. Consequently the
geometry, which is left intact by the YM field, effectively serves
as a ``background'' spacetime which can be chosen somewhat at our
will (as long as it satisfies the vacuum Einstein equation
$R_{\mu\nu}=0$) and here in this work, we take it to be the
gravitational instanton. Loosely speaking, all the typical GI, including
Taub-NUT metric and Eguchi-Hanson solution, possess the same
topology $R\times S^3$ and similar metric structures. Of course in a
stricter sense, their exact topologies can be distinguished, say, by different
Euler numbers and Hirzebruch signatures [4]. Particularly,
in terms of the concise basis 1-forms, the metrics of these GI can
be written as [4]
\begin{eqnarray}
ds^2 &=& c^2_{r}dr^2 +
c^2_{1}\left(\sigma^2_{1}+\sigma^2_{2}\right) +
c^2_{3}\sigma^2_{3} \nonumber \\ &=& c^2_{r}dr^2 +
\sum_{a=1}^{3}c^2_{a}\left(\sigma^{a}\right)^2 = e^{A}\otimes
e^{A}
\end{eqnarray}
where $c_{r}=c_{r}(r)$, $c_{a}=c_{a}(r)$, $c_{1}=c_{2}\neq c_{3}$
and the orthonormal basis 1-form $e^{A}$ is given by
\begin{eqnarray}
e^{A} = \left\{e^{0}=c_{r}dr, ~~e^{a}=c_{a}\sigma^{a}\right\}
\end{eqnarray}
and $\left\{\sigma^{a}\right\}$ ($a=1,2,3$) are the left-invariant
1-forms satisfying the SU(2) Maurer-Cartan structure equation
\begin{eqnarray}
d\sigma^{a} = -{1\over 2}\epsilon^{abc}\sigma^{b}\wedge
\sigma^{c}.
\end{eqnarray}
They form a basis on the $S^{3}$ section of the geometry and hence
can be represented in terms of 3-Euler angles $0\leq \theta
\leq\pi$, $0\leq \phi \leq 2\pi$, and $0\leq \psi \leq 4\pi$
parametrizing $S^3$ as
\begin{eqnarray}
\sigma^1 &=& -\sin\psi d\theta + \cos\psi \sin\theta d\phi, \nonumber \\
\sigma^2 &=&  \cos\psi d\theta + \sin\psi \sin\theta d\phi, \\
\sigma^3 &=& -d\psi - \cos\theta d\phi. \nonumber
\end{eqnarray}
Now in order to construct exact YM instanton solutions in the
background of these GI, we now choose the relevant ans\H{a}tz for
the YM gauge potential and the SU(2) gauge fixing. And in doing
so, our general guideline is that the YM gauge field ans\H{a}tz
should be endowed with the symmetry inherited from that of the
background geometry, the GI. Thus we first ask what kind of
isometry these GI possess. As noted above, typical GI, including
the Taub-NUT and the Eguchi-Hanson metrics, possess the topology of
$R\times S^3$. The geometrical structure of the $S^3$ section,
however, is not that of perfectly ``round'' $S^3$ but rather, that
of ``squashed'' $S^3$. In order to get a closer picture of this
squashed $S^3$, we notice that the $r=$constant slices of these GI
can be viewed as U(1) fibre bundles over $S^2\sim CP^1$ with the
line element
\begin{eqnarray}
d\Omega^2_{3} = c^2_{1}\left(\sigma^2_{1}+\sigma^2_{2}\right) +
c^2_{3}\sigma^2_{3} = c^2_{1}d\Omega^2_{2} +
c^2_{3}\left(d\psi + B\right)^2
\end{eqnarray}
where $d\Omega^2_{2}=(d\theta^2 + \sin^2\theta d\phi^2)$ is the
metric on unit $S^2$, the base manifold whose volume form
$\Omega_{2}$ is given by $\Omega_{2} = dB$ as $B = \cos\theta
d\phi$ and $\psi$ then is the coordinate on the U(1)$\sim S^1$
fibre manifold. Now then the fact that $c_{1}=c_{2}\neq c_{3}$
indicates that the geometry of this fibre bundle manifold is not
that of round $S^3$ but that of squashed $S^3$ with the squashing
factor given by $(c_{3}/c_{1})$. And further, it is squashed along
the U(1) fibre direction. Thus this failure for the geometry to be
that of exactly round $S^3$ keeps us from writing down the
associated ans\H{a}tz for the YM gauge potential right away.
Apparently, if the geometry were that of round $S^3$, one would
write down the YM gauge field ans\H{a}tz as $A^{a}=f(r)\sigma^{a}$
[3] with $\{\sigma^{a}\}$ being the left-invariant 1-forms introduced
earlier. The rationale for this choice can be stated
briefly as follows. First, since the $r=$constant sections of the
background space have the geometry of round $S^3$ and hence
possess the SO(4)-isometry, one would look for the SO(4)-invariant
YM gauge connection ans\H{a}tz as well. Next, noticing that both
the $r=$constant sections of the frame manifold and the SU(2) YM
group manifold possess the geometry of round $S^3$, one may
naturally choose the left-invariant 1-forms $\{\sigma^{a}\}$ as
the ``common'' basis for both manifolds. Thus this YM gauge
connection ans\H{a}tz, $A^{a}=f(r)\sigma^{a}$ can be thought of as
a hedgehog-type ans\H{a}tz where the group-frame index mixing is
realized in a simple manner [3]. Then coming back to our present
interest, namely the GI given in eq.(5), in $r=$constant sections,
the SO(4)-isometry is partially broken down to that of SO(3) by
the squashedness along the U(1) fibre direction to a degree set by
the squashing factor $(c_{3}/c_{1})$. Thus now our task became
clearer and it is how to encode into the YM gauge connection
ans\H{a}tz this particular type of SO(4)-isometry breaking coming
from the squashed $S^3$. Interestingly, a clue to this puzzle can
be drawn from the work of Eguchi and Hanson [7] in which they
constructed abelian instanton solution in Euclidean Taub-NUT
metric (namely the abelian gauge field with (anti)self-dual field
strength with respect to this metric). To get right to the point,
the working ans\H{a}tz they employed for the abelian gauge field
to yield (anti)self-dual field strength is to align the abelian
gauge connection 1-form along the squashed direction, i.e., along
the U(1) fibre direction, $A = g(r)\sigma^3$. This choice looks
quite natural indeed. After all, realizing that embedding of a
gauge field in a geometry with high degree of isometry is itself
an isometry (more precisly isotropy)-breaking action, it would be
natural to put it along the direction in which part of the
isometry is already broken. Finally therefore, putting these two
pieces of observations carefully together, now we are in the
position to suggest the relevant ans\H{a}tz for the YM gauge
connection 1-form in these GI and it is
\begin{eqnarray}
A^{a} = f(r)\sigma^{a} + g(r)\delta^{a3}\sigma^{3}
\end{eqnarray}
which obviously would need no more explanatory comments except
that in this choice of the ans\H{a}tz, it is implicitly understood
that the gauge fixing $A_{r}=0$ is taken. From this point on, the
construction of the YM instanton solutions by solving the
(anti)self-dual equation given in eq.(4) is straightforward. To
sketch briefly the computational algorithm, first we obtain the YM
field strength 2-form (in orthonormal basis) via exterior calculus
(since the YM gauge connection ans\H{a}tz is given in
left-invariant 1-forms) as $F^{a}=(F^{1}, F^{2}, F^{3})$ where
\begin{eqnarray}
F^{1} &=& {f'\over c_{r}c_{1}}(e^0\wedge e^1) + {{f[(f-1)+g]}\over
c_{2}c_{3}}(e^{2}\wedge e^{3}), \nonumber \\
F^{2} &=& {f'\over c_{r}c_{2}}(e^0\wedge e^2) + {{f[(f-1)+g]}\over
c_{3}c_{1}}(e^{3}\wedge e^{1}),  \\
F^{3} &=& {(f'+g')\over c_{r}c_{3}}(e^0\wedge e^3) + {{[f(f-1)-g]}\over
c_{1}c_{2}}(e^{1}\wedge e^{2}) \nonumber
\end{eqnarray}
from which we can read off the (anti)self-dual equation to be
\begin{eqnarray}
\pm {f'\over c_{r}c_{1}} = {{f[(f-1)+g]}\over c_{2}c_{3}},
~~~\pm {(f'+g')\over c_{r}c_{3}} = {{[f(f-1)-g]}\over c_{1}c_{2}}
\end{eqnarray}
where ``$+$'' for self-dual and ``$-$'' for anti-self-dual
equation and we have only a set of two equations as $c_{1}=c_{2}$.
The specifics of different GI are characterized by particular
choices of the orthonormal basis $e^{A} = \{e^{0}=c_{r}dr,
~~e^{a}=c_{a}\sigma^{a}\}$. Thus next, for each GI (i.e., for each
choice of $e^{A}$), we solve the (anti)self-dual equation in (12)
for ans\H{a}tz functions $f(r)$ and $g(r)$ and finally from which
the YM instanton solutions in eq.(10) and their (anti)self-dual
field strength in eq.(11) can be obtained. We now present the
solutions obtained by applying the algorithm presented here to the
two best-known GI, the Taub-NUT and the Eguchi-Hanson metrics.\\
(I) YM instanton in Taub-NUT (TN) metric background \\
The TN GI solution written in the metric form given in eq.(5)
amounts to
\begin{eqnarray}
c_{r}={1\over 2}\left[{r+m\over r-m}\right]^{1/2},
~~~c_{1}=c_{2}={1\over 2}\left[r^2-m^2\right]^{1/2},
~~~c_{3}=m\left[r-m\over r+m\right]^{1/2} \nonumber
\end{eqnarray}
and it is a solution to Euclidean vacuum Einstein equation
$R_{\mu\nu}=0$ for $r\geq m$ with self-dual Riemann tensor. The
apparent singularity at $r=m$ can be removed by a coordinate
redefinition and is a `nut' (in terminology of Gibbons and Hawking
[4]) at which the isometry generated by the Killing vector
$(\partial/\partial \psi)$ has a zero-dimensional fixed point set.
And this TN instanton is an asymptotically-locally-flat (ALF) metric. \\
It turns out that only the anti-self-dual equation
$F^{a}=-\tilde{F}^{a}$ admits a non-trivial solution and it is
$A^{a}=(A^1, A^2, A^3)$ where
\begin{eqnarray}
A^1 = \pm 2{(r-m)^{1/2}\over (r+m)^{3/2}}e^1,  ~~~A^2 = \pm
2{(r-m)^{1/2}\over (r+m)^{3/2}}e^2, ~~~A^3 = {(r+3m)\over m}
{(r-m)^{1/2}\over (r+m)^{3/2}}e^3
\end{eqnarray}
and $F^{a}=(F^1, F^2, F^3)$ where
\begin{eqnarray}
F^1 &=& \pm {8m\over (r+m)^3}\left(e^0\wedge e^1 - e^2\wedge
e^3\right), ~~~F^2 = \pm {8m\over (r+m)^3}\left(e^0\wedge e^2 -
e^3\wedge e^1\right), \nonumber \\ F^3 &=&  {16m\over
(r+m)^3}\left(e^0\wedge e^3 - e^1\wedge e^2\right).
\end{eqnarray}
It is interesting to note that this YM field strength and the
Ricci tensor of the background TN GI are proportional as
$|F^{a}|=2|R^{0}_{a}|$ except for opposite self-duality, i.e.,
\begin{eqnarray}
R^0_{1}=-R^2_3 &=&  {4m\over (r+m)^3}\left(e^0\wedge e^1 +
e^2\wedge e^3\right), ~~~R^0_{2}=-R^3_1 = {4m\over
(r+m)^3}\left(e^0\wedge e^2 + e^3\wedge e^1\right), \nonumber \\
R^0_{3}=-R^1_2 &=& -{8m\over (r+m)^3}\left(e^0\wedge e^3 +
e^1\wedge e^2\right).
\end{eqnarray}
(II) YM instanton in Eguchi-Hanson (EH) metric background \\ The
EH GI solution amounts to
\begin{eqnarray}
c_{r}=\left[1 - \left({a\over r}\right)^4\right]^{-1/2},
~~~c_{1}=c_{2}={1\over 2}r, ~~~c_{3}={1\over 2}r\left[1 - \left({a\over
r}\right)^4 \right]^{1/2} \nonumber
\end{eqnarray}
and again it is a solution to Euclidean vacuum Einstein equation
$R_{\mu\nu}=0$ for $r\geq a$ with self-dual Riemann tensor. $r=a$
is just a coordinate singularity that can be removed by a
coordinate redefinition provided that now $\psi$ is identified
with period $2\pi$ rather than $4\pi$ and is a `bolt' (in
terminology of Gibbons and Hawking [4]) where the action of the
Killing field $(\partial/\partial \psi)$ has a two-dimensional
fixed point set. Besides, this EH instanton is an
asymptotically-locally-Euclidean (ALE) metric. \\
In this time, only the
self-dual equation $F^{a}=+\tilde{F}^{a}$ admits a non-trivial
solution and it is $A^{a}=(A^1, A^2, A^3)$ where
\begin{eqnarray}
A^1 = \pm {2\over r}\left[1 - \left({a\over r}\right)^4\right]^{1/2}e^1,
~~~A^2 = \pm {2\over r}\left[1 - \left({a\over r}\right)^4\right]^{1/2}e^2,
~~~A^3 = {2\over r}{\left[1 + \left({a\over r}\right)^4\right]\over \left[1 -
\left({a\over r}\right)^4\right]^{1/2}} e^3
\end{eqnarray}
and $F^{a}=(F^1, F^2, F^3)$ where
\begin{eqnarray}
F^1 &=& \pm {4\over r^2}\left({a\over r}\right)^4\left(e^0\wedge e^1 +
e^2\wedge e^3\right), ~~~F^2 = \pm {4\over r^2}\left({a\over
r}\right)^4\left(e^0\wedge e^2 + e^3\wedge e^1\right), \nonumber \\ F^3
&=& - {8\over r^2}\left({a\over r}\right)^4\left(e^0\wedge e^3 + e^1\wedge
e^2\right).
\end{eqnarray}
Again it is interesting to realize that this YM field strength and
the Ricci tensor of the background EH GI are proportional as
$|F^{a}|=2|R^{0}_{a}|$, i.e.,
\begin{eqnarray}
R^0_{1}=-R^2_3 &=& {2\over r^2}\left({a\over r}\right)^4\left(- e^0\wedge e^1
+ e^2\wedge e^3\right), ~~~R^0_{2}=-R^3_1 = {2\over r^2}\left({a\over
r}\right)^4\left(- e^0\wedge e^2 + e^3\wedge e^1\right), \nonumber \\
R^0_{3}=-R^1_2 &=& - {4\over r^2}\left({a\over r}\right)^4\left(- e^0\wedge
e^3 + e^1\wedge e^2\right).
\end{eqnarray}
It is also interesting to note that this YM instanton solution particularly
in EH background (which is ALE) obtained by directly solving the self-dual
equation can also be ``constructed'' by simply identifying
$A^{a}=\pm 2\omega^{0}_{a}$ (where $\omega^{0}_{a}=(\epsilon_{abc}/2)\omega^{bc}$
are the spin connection of EH metric) and hence $F^{a}=\pm 2R^{0}_{a}$ as was
noticed by [10] but in the string theory context with different motivation.
This construction of solution via a simple identification of gauge field
connection with the spin connection, however, works only in ALE backgrounds
such as EH metric and generally fails as is manifest in the previous TN
background case (which is ALF, not ALE) in which $A^{a}\neq \pm 2\omega^{0}_{a}$
but still $F^{a}=\pm 2R^{0}_{a}$. Thus the method
presented here by first writing (by employing a relevant ans\H{a}tz for the YM gauge
connection given in eq.(10)) and directly solving the (anti) self-dual equation
looks to be the algorithm for generating the solution with general applicability
to all species of GI in a secure and straightforward manner. Indeed, the detailed
and comprehensive coverage of YM instanton solutions in all other GI based
on the algorithm presented in this work will be reported elsewhere and it will
show how simple albeit powerful this method really is. In this regard, the
method for generating YM instanton solutions to (anti) self-dual equation in
all known GI backgrounds proposed here in this work can be contrasted to
earlier works in the literature [12] discussing the construction of YM instantons
mainly in the background of ALE GI via indirect methods such as that of ADHM [11]. \\
Having constructed explicit YM instanton solutions in TN and EH
GI, we now turn to the physical interpretation of the structure of
these SU(2) YM instantons supported by the two typical GI. Recall
that the relevant ans\H{a}tz for the YM gauge connection is of the
form $A^{a}=f(r)\sigma^{a}$ in the background geometry such as de
Sitter GI [3] with topology of $R\times ({\rm round})S^3$ and of the
form $A^{a}=f(r)\sigma^{a} + g(r)\delta^{a3}\sigma^3$ in the less
symmetric GI backgrounds with topology of $R\times ({\rm
squashed})S^3$. Thus in order to get some insight into the
physical meaning of the structure of these YM connection
ans\H{a}tz, we now try to re-express the left-invariant 1-forms
$\{\sigma^{a}\}$ forming a basis on $S^3$ in terms of more
familiar Cartesian coordinate basis. Utilizing the coordinate
transformation from polar $(r, ~\theta, ~\phi, ~\psi)$ to
Cartesian $(t, x,y,z)$ coordinates (note, here, that $t$ is not
the usual ``time'' but just another spacelike coordinate) given by
[4]
\begin{eqnarray}
x+iy = r\cos {\theta\over 2}e^{{i\over 2}(\psi+\phi)}, ~~~z+it =
r\sin {\theta\over 2}e^{{i\over 2}(\psi-\phi)},
\end{eqnarray}
where $x^2+y^2+z^2+t^2=r^2$ and further introducing
the so-called `tHooft tensor [1,9] defined by
$\eta^{a\mu\nu}=-\eta^{a\nu\mu}=(\epsilon^{0a\mu\nu}+\epsilon^{abc}
\epsilon^{bc\mu\nu}/2)$, the left-invariant 1-forms can be cast to
a more concise form
$\sigma^{a}=2\eta^{a}_{\mu\nu}(x^{\nu}/r^2)dx^{\mu}$. Therefore,
the YM instanton solution, in Cartesian coordinate basis, can be
written as
\begin{eqnarray}
A^{a} = A^{a}_{\mu}dx^{\mu} =
2\left[f(r)+g(r)\delta^{a3}\right]\eta^{a}_{\mu\nu}{x^{\nu}\over
r^2}dx^{\mu}
\end{eqnarray}
in the background of TN and EH GI with topology of $R\times ({\rm
squashed})S^3$. Now to appreciate the meaning of this structure,
we go back to the flat space situation. As is well-known, in flat
space, the standard BPST [1] SU(2) YM instanton solution takes the
form $A^{a}_{\mu} = 2\eta^{a}_{\mu\nu}[x^{\nu}/(r^2+\lambda^2)]$
with $\lambda$ being the size of the instanton. Note, however,
that separately from this BPST instanton solution, there is
another non-trivial solution to the YM field equation of
the form $A^{a}_{\mu} = \eta^{a}_{\mu\nu}(x^{\nu}/r^2)$ found long
ago by De Alfaro, Fubini, and Furlan [8]. This second solution is
called ``meron'' [9] as it carries a half unit of topological charge
and is known to play a certain role
concerning the quark confinement [9]. It, however, exhibits
singularity at its center $r=0$ and hence has a diverging action and falls
like $1/r$ as $r\rightarrow \infty$. Thus we are led to the conclusion
that the YM instanton solution in typical GI backgrounds possess the
structure of (curved space version of) meron at large $r$. As is well-known,
in flat spacetime meron does not solve the 1st order (anti) self-dual
equation although it does the second order YM field equation.
Thus in this sense, this result seems remarkable since it implies
that in the GI backgrounds, the (anti) self-dual YM equation admits
solutions which exhibit the configuration of meron solution at large $r$
in contrast to the flat spacetime case. And we only
conjecture that when passing from the flat ($R^4$) to GI ($R\times
S^3$) geometry, the closure of the topology of part of the
manifold appears to turn the structure of the instanton solution
from that of standard BPST into that of meron. Next, we look into
the behavior of these solutions in TN and EH GI
backgrounds as $r\rightarrow 0$. For TN and EH instantons, the
ranges for radial coordinates are $m\leq r <\infty$ and $a\leq r
<\infty$, respectively. Since the point $r=0$ is absent in these
manifolds, the solutions in these GI are everywhere
regular. Finally, we close with perhaps the most interesting
comments on the estimate of the instanton contribution to the
intervacua tunnelling amplitude. It has been pointed out in the
literature that both in the background of Euclidean Schwarzschild
geometry [2] and in the Euclidean de Sitter space [3], the (anti)
instanton solutions have the Pontryagin index of $\nu[A]= \pm 1$
and hence give the contribution to the (saddle point approximation
to) intervacua tunnelling amplitude of $\exp{[-8\pi^2/g^2_{c}]}$,
which, interestingly, are the same as their flat space
counterparts even though these curved space YM instanton solutions
do not correspond to gauge transformations of any flat space
instanton solution [1]. This unexpected and hence rather curious
property, however, turns out not to persist in YM instantons
in GI backgrounds such as TN and EH
metrics. In order to see this, consider the curved space version
of Pontryagin index or second Chern class having the
interpretation of instanton number $\nu[A]$ given by
\begin{eqnarray}
\nu[A] = Ch_{2}(F) = {-1\over 8\pi^2}\int_{M^4}tr(F\wedge F) =
\int_{R\times S^3}d^4x\sqrt{g}\left[{-1\over
32\pi^2}F^{a}_{\mu\nu}\tilde{F}^{a\mu\nu}\right]
\end{eqnarray}
and the saddle point approximation to the intervacua tunnelling
amplitude
\begin{eqnarray}
\Gamma_{GI} \sim \exp{[-I_{GI}(instanton)]}
\end{eqnarray}
where the subscript ``GI'' denotes corresponding quantities in
the GI backgrounds and $I_{GI}(instanton)$ represents the
Euclidean YM theory action evaluated at the YM instanton solution,
i.e.,
\begin{eqnarray}
I_{GI}(instanton) = \int_{R\times S^3}d^4x\sqrt{g}\left[{1\over
4g^2_{c}}F^{a}_{\mu\nu}F^{a\mu\nu}\right] = \left({8\pi^2\over
g^2_{c}}\right)|\nu[A]|
\end{eqnarray}
where we used the (anti)self-duality relation $F^{a}=\pm
\tilde{F}^{a}$. Then the straightforward calculation yields ;
$\nu[A] = 1$, $I_{GI}(instanton)=8\pi^2/g^2_{c}$ and $\Gamma_{GI}
\sim \exp{(-8\pi^2/g^2_{c})}$ for the instanton solution in TN
metric and $\nu[A] = -3/2$, $I_{GI}(instanton)=12\pi^2/g^2_{c}$
and $\Gamma_{GI} \sim \exp{(-12\pi^2/g^2_{c})}$ for the instanton
solution in EH metric background. Here, however, the solution in
EH metric background carries the half-integer Pontryagin index
actually because the boundary of EH space is $S^3/Z_2$ [13].
Therefore we need to be cautious in drawing the conclusion that
the fact that solutions in GI backgrounds carry fractional
topological charges could be another supporting evidence for meron
interpretation of the solutions. To summarize, in the present work
we constructed the solutions to (anti)self-dual YM equation in the
typical gravitational instanton geometries and analyzed their
physical nature. As demonstrated, the solutions turn out to take
the structure of merons at large $r$ and generally carry
fractional topological charge values. Nevertheless, it seems more
appropriate to conclude that the solutions still should be
identified with (curved space version of) instantons as they are
solutions to 1st order (anti) self-dual equation and are
everywhere regular having finite YM action. However, these curious
mixed characteristics of the solutions to (anti) self-dual YM
equation in GI backgrounds appear to invite us to take them more
seriously and further explore potentially interesting physics
associated with them. \\ This work was supported in part by BK21
project in physics department at Hanyang Univ. and by grant No.
1999-2-112-003-5 from the interdisciplinary research program of
the KOSEF.

\vspace*{2cm}

\noindent

\begin{center}
{\rm\bf References}
\end{center}

\begin{description}

\item {[1]} A. A. Belavin, A. M. Polyakov, A. S. Schwarz, and Yu. S. Tyupkin,
            Phys. Lett. {\bf B59}, 85 (1975) ;
            G. `tHooft, Phys. Rev. Lett. {\bf 37}, 8 (1976).
\item {[2]} J. M. Charap and M. J. Duff, Phys. Lett. {\bf B69}, 445 (1977) ;
            {\it ibid} {\bf B71}, 219 (1977).
\item {[3]} H. Kim and S. K. Kim,
            Nuovo Cim. {\bf B114}, 207 (1999) and references therein.
\item {[4]} T. Eguchi, P. B. Gilkey, and A. J. Hanson, Phys. Rep. {\bf 66}, 213 (1980) ;
            G. W. Gibbons and C. N. Pope, Commun. Math. Phys. {\bf 66}, 267 (1979) ;
            G. W. Gibbons and S. W. Hawking, {\it ibid}, {\bf 66}, 291 (1979).
\item {[5]} A. Taub, Ann. Math. {\bf 53}, 472 (1951) ;
            E. Newman, L. Tamburino, and T. Unti, J. Math. Phys. {\bf 4}, 915 (1963) ;
            S. W. Hawking, Phys. Lett. {\bf A60}, 81 (1977).
\item {[6]} T. Eguchi and A. J.Hanson, Phys. Lett. {\bf B74}, 249 (1978).
\item {[7]} T. Eguchi and A. J. Hanson, Ann Phys. {\bf 120}, 82 (1979).
\item {[8]} V. De Alfaro, S. Fubini, and G. Furlan, Phys. Lett. {\bf B65}, 163 (1976).
\item {[9]} C. G. Callan, R. Dashen, and D. J. Gross, Phys. Rev. {\bf D17}, 2717 (1978).
\item {[10]} M. Bianchi, F. Fucito, G. C. Rossi, and M. Martellini, Nucl. Phys. {\bf B440},
             129 (1995).
\item {[11]} M. Atiyah, V. Drinfeld, N. Hitchin, and Y. Manin, Phys. Lett. {\bf A65}, 185
             (1987) ; P. B. Kronheimer, and H. Nakajima, Math. Ann. {\bf 288}, 263 (1990).
\item {[12]} H. Boutaleb-Joutei, A. Chakrabarti, and A. Comtet, Phys. Rev. {\bf D20}, 1844
             (1979) ; {\it ibid.} {\bf D20}, 1898 (1979) ; {\it ibid.} {\bf D21}, 979 (1980) ;
             {\it ibid.} {\bf D21}, 2280 (1980) ; A. Chakrabarti, Fortschr. Phys. {\bf 35}, 1
             (1987) ;  M. Bianchi, F. Fucito, G. C. Rossi, and M. Martellini, Phys. Lett.
             {\bf B359}, 49-61 (1995).
\item {[13]} We thank the anonymous referee for pointing this out to us.

\end{description}

\end{document}